\documentclass[twocolumn,superscriptaddress,amsmath,amssymb,aps,prl]{revtex4-2}
\usepackage[utf8]{inputenc}

\usepackage[]{siunitx}
\usepackage{amsmath}
\usepackage{amsfonts}
\usepackage{amsthm}
\usepackage{mathtools} % for dcases
\usepackage{acronym}
\usepackage[hyperindex,breaklinks]{hyperref}
\usepackage{comment}
\usepackage{gensymb}
\usepackage{graphicx}
\usepackage{dcolumn}
\usepackage{array}
\usepackage{booktabs}
\usepackage{bm}% bold math
\usepackage{upgreek}
\usepackage{xcolor}
\newcommand{\vcr}[1]{\boldsymbol{\mathrm{#1}}}

\newcommand{\braket}[1]{\left\langle #1 \right\rangle}

\newcommand{\Abs}[1]{\left\vert #1 \right\vert}

\newcommand{\SqB}[1]{ \left[ #1 \right]}
\newcommand{\RnB}[1]{ \left( #1 \right)}
\newcommand{\CrB}[1]{ \left\{ #1 \right\}}

\newcommand{\myFigWide}[6]{ %
\begin{figure*}[htb]
\begin{center}
\includegraphics[width=#1\columnwidth,height=#2\columnwidth,clip=true,keepaspectratio=#3]{#4}
\caption{#5} \vspace{-0.5cm} \label{#6} 
\end{center}
\end{figure*}}

\begin{document}
\title{Phonon and magnon jets above the critical current in nanowires with planar domain walls}
\author{Maria Stamenova}\email[Contact email address: ]{stamenom@tcd.ie}
\affiliation{School of Physics and CRANN, Trinity College, Dublin 2, Ireland} 
\author{Plamen Stamenov}
\affiliation{School of Physics and CRANN, Trinity College, Dublin 2, Ireland}
\author{Tchavdar Todorov}
\affiliation{Centre for Quantum Materials and Technologies, Queen’s University Belfast, Belfast, UK}

\begin{abstract}
We show through non-equilibrium non-adiabatic electron-spin-lattice simulations that above a critical current in magnetic atomic wires with a narrow domain wall (DW), a couple of atomic spaces in width, the electron flow triggers violent stimulated emission of phonons and magnons with an almost complete conversion of the incident electron momentum flux into a phonon and magnon flux. Just below the critical levels of the current flow, the DW achieves maximal velocity of about $3\times 10^{4}$ m/s, entering a strongly non-adiabatic regime of DW propagation, followed by a breakdown at higher biases. Above this threshold a further increase of the current with the applied bias is impossible -- the electronic current suffers a heavy suppression and the DW stops. This poses a fundamental limit to the current densities attainable in atomic wires. At the same time it opens up an exciting way of generating the alternative quasi-particle currents, described above, once the requisite electronic-structure properties are met. 
\end{abstract}


\maketitle
\section {Introduction} \label{sect:intro}

Electron-phonon scattering is how electrons and nuclei in solids maintain thermal equilibrium. Nanoscale devices allow enormous current densities leading to violent electron-phonon dynamics. The most familiar is Joule heating, driven by spontaneous phonon emission \cite{tnt1,mads1}. For a long time it was considered a central stability-limiting mechanism for small conductors. However, it is now becoming clear that stimulated processes can play an even bigger role. Two key effects, that can be traced to these processes, are electronic friction \cite{langevin} and non-conservative current-induced forces and torques \cite{bjnano,bulk,MS05}.

Current-induced magnetic domain wall (DW) motion due to spin-transfer torques (STTs), but also the interaction between existing magnetic texture and excited spin-waves (SWs), which we choose to call {\it magnons} here, is instrumental for new spintronic logic and data-storage technologies, also underpinning the emerging field of magnonics. For instance, fast-moving DWs can be sources of strong tunable SWs \cite{Yan2011}, but DWs can also act as waveguides\cite{Garcia-Sanchez2015, DWguides21} for SWs, transmitting information without losses \cite{Wagner2016, Garcia-Sanchez2015, DWguides21}. Conversely, magnonic currents transfer angular momentum to DWs, similarly to STTs, resulting in DW translational motion \cite{Wang12}. But they can also give rise to magnonic drag or magnonic barrier, slowing the DW motion \cite{Tatara08,Yan2011}. The high-velocity motion of DWs is, in principle, limited by the non-adiabatic excitation of low- and medium-wavevector ($q$) magnons, which is otherwise suppressed by the anisotropy gap in their dispersion.\cite{Baryaktar1994} 

Here we show, through time-dependent open-boundary simulations coupling electronic, lattice and localised-spin degrees of freedom, that stimulated phonon emission poses a fundamental limit on the currents that can be passed with impunity through electronic and spintronic devices. There is a critical current density of $10^{12} - 10^{13}$ A${\rm m}^{-2}$, above which the current generates violent phonon jets with an almost complete conversion of the electron momentum flux into phonon momentum flux. A related phenomenon occurs in magnetic structures with sharp non-collinear features -- the emission of magnon jets as a result of the exchange interaction between the itinerant electrons and the DW spins. In both cases the emitted waves are nearly monochromatic with momentum $2k_\mathrm{F}$ for phonons and $|k_\mathrm{F}^{\uparrow}-k_\mathrm{F}^{\downarrow}|$ for magnons. Because these processes are stimulated, they are captured by the simplest form of non-adiabatic dynamics, Ehrenfest dynamics, which we have implemented in our open-boundary electron-phonon-spin simulatons.

The origin of these processes is uncompensated stimulated directional quasi-particle emission, discussed in the Supplemental Material (SM) \cite{SM} (see also references \onlinecite{ejp,eunan,valerio,kosov,mads_laser,valerio,waterwheel,berry,mads_bulk,pq,Pertsova2011} therein) for phonons. When the energy window for conduction $eV_b$ exceeds $\hbar\omega$, electrons can emit forward-travelling phonons and backscatter into a lower-energy state, while there are no electrons in the conduction window travelling the other way to reabsorb these phonons. The result is uncompensated stimulated forward phonon emission -- the process we are interested in. The condition $eV_b = \hbar\omega$ is equivalent to $v_{\rm drift} = c$ in 1D:
\begin{equation}
v_{\rm drift} = \frac{eV_b}{\pi\hbar}\frac{\pi}{2k_\mathrm{F}} = \frac{\omega}{q} = c
\end{equation}
where we use quasi-momentum conservation $q \sim 2k_\mathrm{F}$ and assume dispersionless phonons for simplicity. The critical bias for the process $eV_b \sim \hbar\omega$ is between 10s and 100s of meV, well in the nano-electronics experimental range. The corresponding current densities are $10^{12} - 10^{13}$ A${\rm m}^{-2}$.

\myFigWide{2}{1}{true}{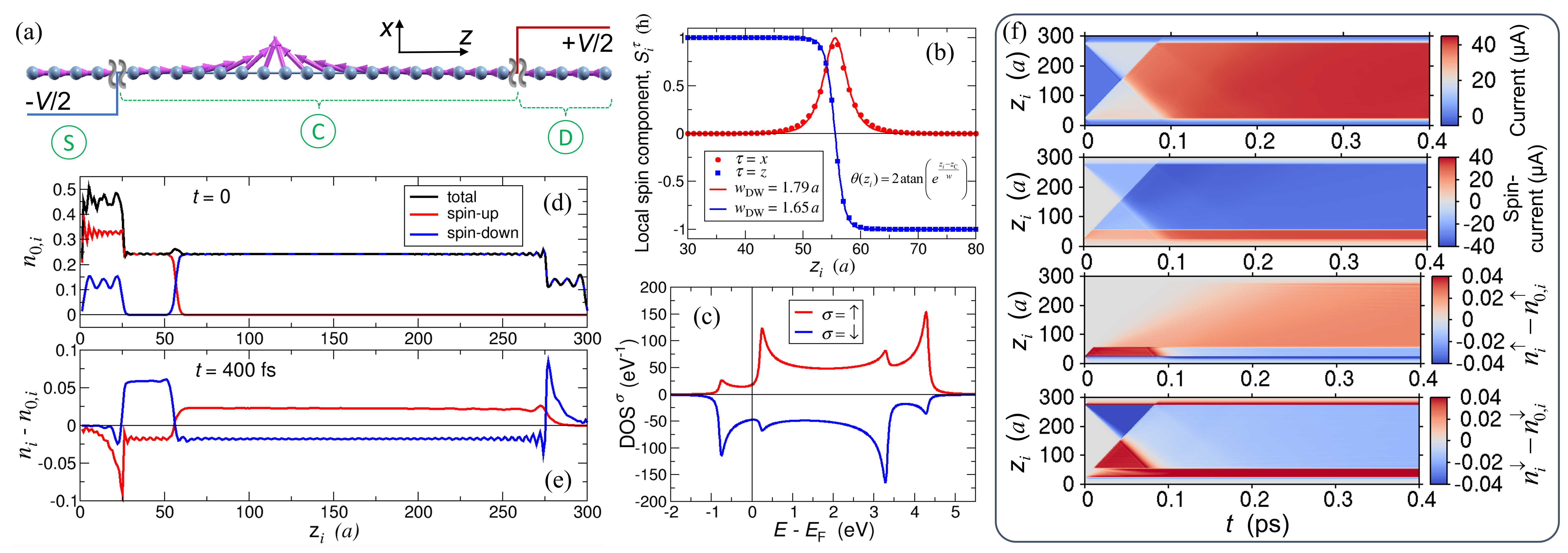}{(a) Schematic of the relaxed planar DW formed by the local spins at $t=0$, showing the potential drop applied to the ends of the chain ($V_b>0$). (b) In-plane components of the local spins at $t=0$ and fits to the analytical profile from the 1D anisotropic Heisenberg model. (c) DOS for the self-consistent ground state of uniform wire with a relaxed DW at $t=0$, spin-polarisation is defined wrt the on-site quantisation axis ($\parallel \vcr{S}_i$). (d,e) Spin-density distribution $n_i=\braket{\sigma^z}_i$ at $t=0$ and its variation wrt $n_{0,i}=n_i(t=0)$ after $t=400$~fs, respectively. (f) Heatmaps over space and time of indicated early-dynamic properties at $V_b=1$~V: charge and spin-polarized currents; variation of each spin-density wrt the $t=0$ distribution [panel (d)].}{fig01}

The prediction is that above these current densities something catastrophic should happen in the atomic dynamics. This is the effect we set out to simulate, together with its analogue for magnons. 

\section{Methodology}

Our system is a 1D atomic wire of $N=300$ atoms with localised spins [Fig. \ref{fig01}(a)]. The itinerant ($s$-shell) electrons are described through a non-collinear spin-dependent tight-binding Hamiltonian with a single real-space basis state $|i,s\rangle$ per atom $i$, for the two possible values of the spin label $s=\pm 1$, 
\begin{eqnarray}  \label{H_e}
H_e &=& \sum_{i,j,s} \big[ \RnB{E_0 + U_\mathrm{C} \Delta n_i}\delta_{ij}  
+ t_{ij}(1-\delta_{ij}) \big] c_{is}^\dagger c_{js} \nonumber \\ 
&-& J_{sd} \sum_{i,s,s'} \RnB{ \vcr{\sigma}_{ss'}\cdot \vcr{S}_i } c_{is}^\dagger c_{is'}  \, ,
\end{eqnarray}  
where $\Delta n_i = n_i - n_{0}$ is the excess onsite charge and $2\vcr{\sigma}$ is the vector of Pauli spin matrices. We explore a half-metallic regime with uniform $n_0=0.25$~$e$/site [see Fig. \ref{fig01}(c)]. Source (S) and drain (D) regions are 25-atom-long segments at each chain end. The central region (C) contains the remaining 250 atoms and the Coulomb interaction $U_\mathrm{C}$ is applied therein. Current-carrying conditions are imposed through a semi-empirical TD open-boundary method\cite{Sanchez06}, based on a modified quantum Liouville equation for the electronic density matrix (see SM\cite{SM}).

The classical degrees of freedom are a set of localised  ($d$-shell) spins $\CrB{\vcr{S}_i}$, one per site, and atomic positions $\CrB{\vcr{R}_i}$. They are both propagated dynamically only in the C region, while in S/D they remain fixed.  For the classical spins we integrate the Liouville equation $\dot{\vcr{S}_i}=\RnB{\vcr{S}_i \times \vcr{B}_i}/\hbar$, where
\begin{equation} \label{LLGel1}
\vcr{B}_i = J_{sd} \braket{\vcr{\sigma}}_i +  J_{dd}\sum_{j \ne i} \eta(R_{ij}) \vcr{S}_j + 2 J_z \RnB{\vcr{S}_i \cdot \hat{\vcr{z}}} \hat{\vcr{z}}
\end{equation} 
is the effective time-dependent magnetic field at site $i$, $\braket{\vcr{\sigma}}_i = \sum_{s,s'}\langle {\hat c}^\dagger_{is}{\hat c}_{is'} \rangle\vcr{\sigma}_{ss'}$, $R_{ij}=\Abs{\vcr{R}_i-\vcr{R}_j}$. All spins $\Abs{{\bf S}_i}=S=1$ are locally exchange-coupled to the instantaneous onsite expectation value of the itinerant electron spin, which gives rise to STT. $\CrB{\vcr{S}_i}$ are also coupled through a distance-dependent Heisenberg interaction with a strength $J_{dd}$ at the nearest-neighbour distance $a$, decaying as $\eta(R_{ij})=\exp{\SqB{-(R_{ij}-a)/a}}$, and $J_z$ is an easy-axis anisotropy. The initial ($t=0$) configuration is obtained through a damped relaxation of $\CrB{\vcr{S}_i}$, as a result of which a stationary narrow planar (N\'eel) DW\cite{Tatara08} is formed along the easy $z$-axis [Fig. \ref{fig01}(a)].  

At $t=0$ the chain is uniform with lattice spacing $a$. Atomic positions in C evolve according to
\begin{equation} \label{F_i}
M \ddot{\vcr{R}}_i = - 2\sum_{j\neq i} \Re\RnB{\rho_{ij}} {\nabla}_{i} t_{ij} - \sum_{j\neq i} {\nabla}_{i} \SqB{ \Phi_{ij} - V_{\mathrm{S},ij} },
\end{equation}
where the hopping parameter $t_{ij}$ and the pair potential $\Phi_{ij}$ are inverse power-law distance-dependent (see SM\cite{SM}) and the local spin interaction $V_{\mathrm{S},ij}=-J_{dd}\,\eta(R_{ij})\,\vcr{S}_i \cdot \vcr{S}_j $. The main parameters are: $J_{sd}=U_\mathrm{C}=1$~eV, $J_{dd}=(20-200)$~meV, $J_z=J_{dd}/2$, $a=2.5$~\AA, $M=10$~amu (see SM\cite{SM} for full list). Our parameters produce compressed chains \cite{long_wires}, in order to suppress a Peierls distortion \cite{long_wires}.

\myFigWide{1.9}{1}{true}{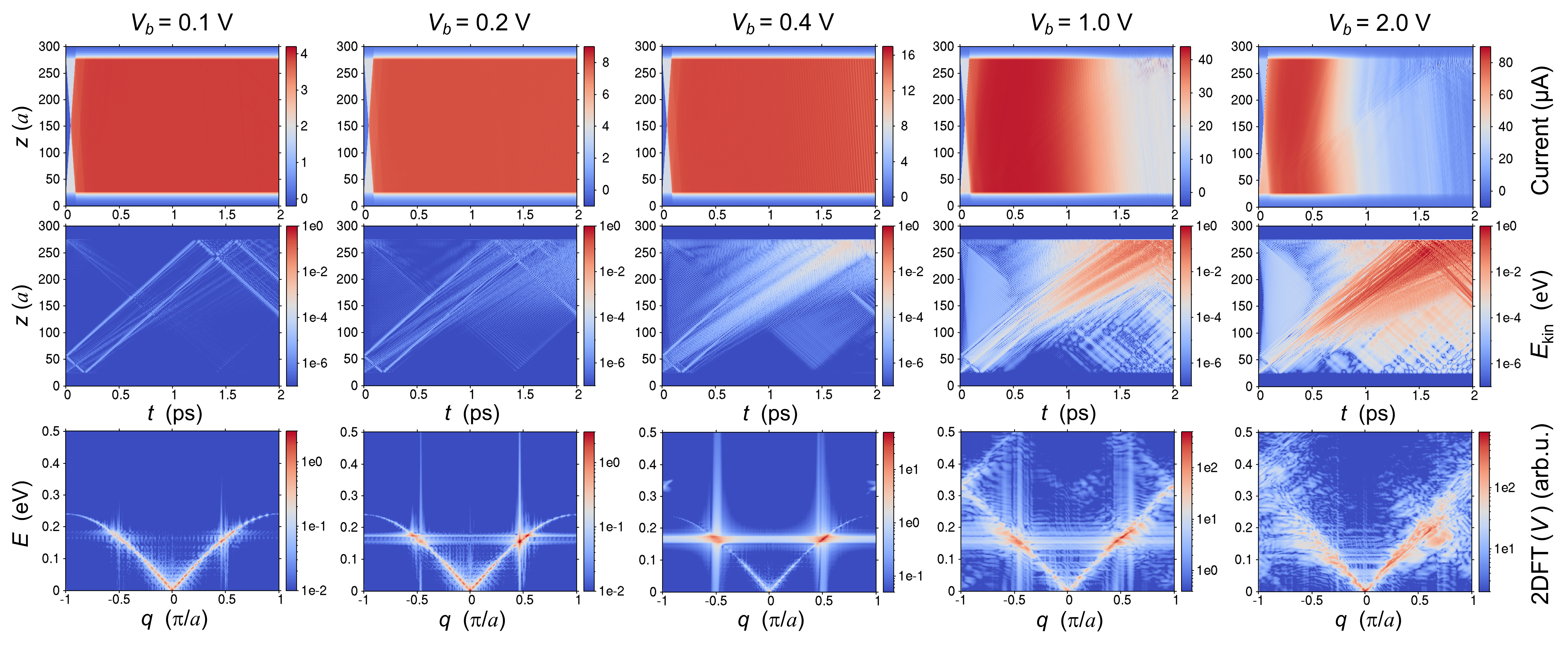}{Dynamics with frozen spins: heatmap plots in columns for given applied bias voltage (between 0.1 and 2 V). The first two rows depict the real-space and time evolution of the bond currents (in $\upmu$A) and the kinetic energy of the atoms (in eV), respectively. In the bottom row are the corresponding 2D-Fourier-transform images of the atomic velocities.}{fig02}

These equations of motion constitute Ehrenfest dynamics for our electron-phonon-magnon problem. Ehrenfest dynamics captures electronic friction, non-conservative forces and Berry forces in transport \cite{ejp} but excludes electronic noise and spontaneous emission for phonons or spin noise for the magnetic degrees of freedom \cite{jpcm_review}. Our interest here is precisely in the stimulated processes that the mean-field Ehrenfest approximation is designed for.

The resultant electronic structure is shown in Fig. \ref{fig01}. The initial position of the DW (around atom \#56) can be seen from the steps in the two spin populations in panels (d, e). Panel (c) shows that for the present parameters the system is half-metallic. Indeed, in panel (f) the evolution of the charge ($I_i^\uparrow + I_i^\downarrow$) and the spin-polarised ($I_i^\uparrow - I_i^\downarrow$) bond currents \cite{bond_currents}, at 1~V of bias (with frozen atoms and local spins), shows average current of $39\,\upmu$A\,V$^{-1}$ (25.6~k$\Omega$), corresponding to one spin channel.

\section{Results and discussion}

We first apply a series of biases with moving ions and frozen magnetic moments, Fig. \ref{fig02}. At about 0.4~V we see two dramatic blasts in the 2D-Fourier transform (2DFT) of the atomic velocities, while the atomic kinetic energies clearly display the phonon jets. They first appear around 0.2~V. The phonon dispersion on the 2DFT plot shows that at a phonon wavevector $q = 2k_\mathrm{F} \approx \pm \pi/2a$ the phonon frequency is in the region of 0.17~eV, corresponding to the activation bias.  In the region $\Abs{q}>\pi/2a$ a linear branch with $E\sim q$ splits off from the regular phonon dispersion and becomes dominant at higher bias (see also SM\cite{SM}). We attribute this to the electron-phonon interaction and the electronic open boundaries.     

\myFigWide{1.9}{1}{true}{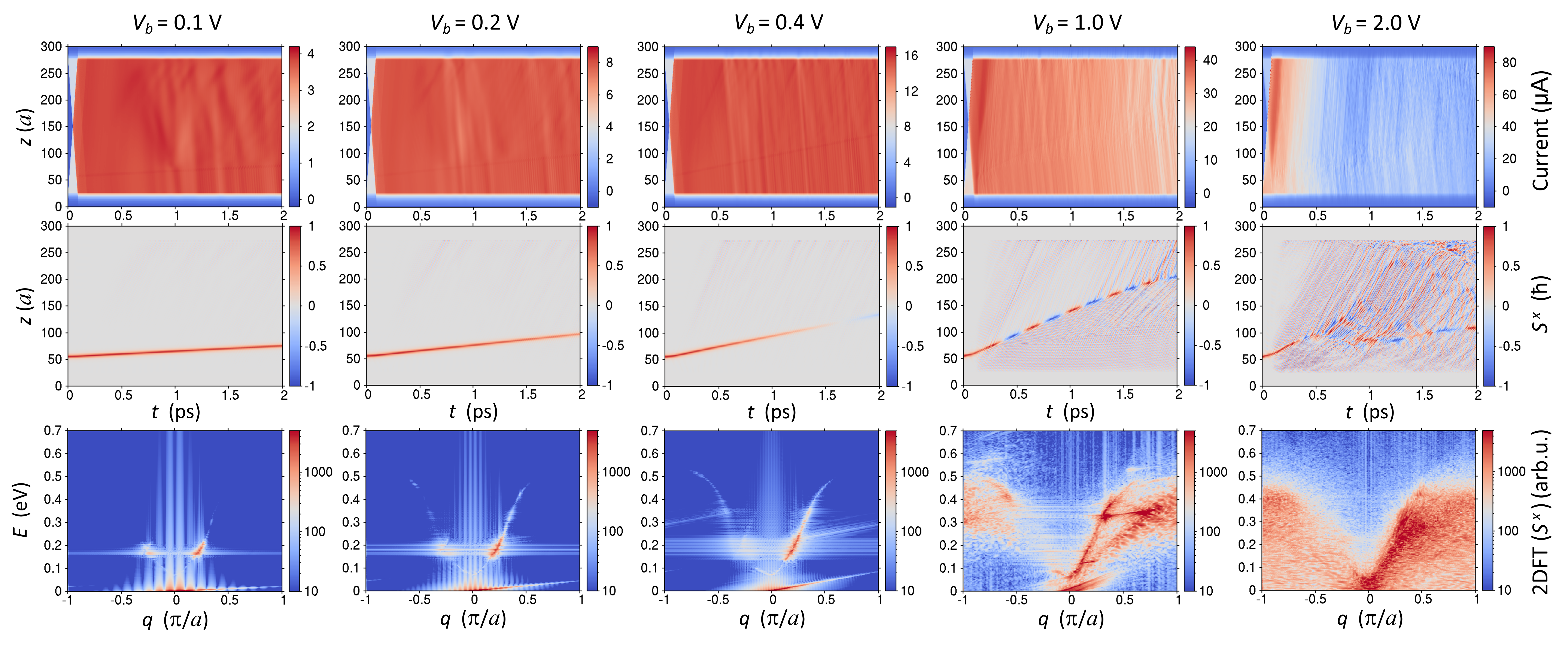}{Dynamics with frozen atoms: heatmap plots in analogy with Fig. \ref{fig02}, here the rows from the top are the current (in $\upmu$A), the $S^z$ component (in $\hbar$) and the 2DFT($S^x$). Here $J_{dd}=80$~meV, $J_z=40$~meV. }{fig03}

These results demonstrate the effect: currents above the threshold generate explosive phonon fluxes and the current drops heavily. Quasi-momentum balance requires the kinetic energy per atom, at the right end, to be $\sim n_0 eV_b / 2$ where $n_0=0.25$ is the number of electrons per atom. This gives a kinetic energy of about 0.13 eV at 1 V, in rough agreement with Fig. \ref{fig02}. Once the phonon jet has been unleashed, the current drops to about the threshold value, around 10 $\upmu$A here, as is seen from Fig. \ref{fig02} at 1~V and 2~V. Notice also the approximately exponential growth of the phonon jet along its propagation. This is the tell-tale sign of the stimulated process, in stark contrast to Joule heating.

We now consider the spin dynamics. Fig. \ref{fig03} shows the magnetisation dynamics with frozen atoms at different biases. At small bias, the spin-polarised current drives DW motion. Notice the signature of the linear motion of the DW on the Fourier spectrum: the linear dispersion, most clearly seen at low bias. With increasing bias the quadratic (for low-$q$) magnon dispersion becomes visible with its lopsided population towards right-moving magnons. However, at and above 1~V the current drops while generating a dramatic magnonic population.

\myFigWide{2.15}{1}{true}{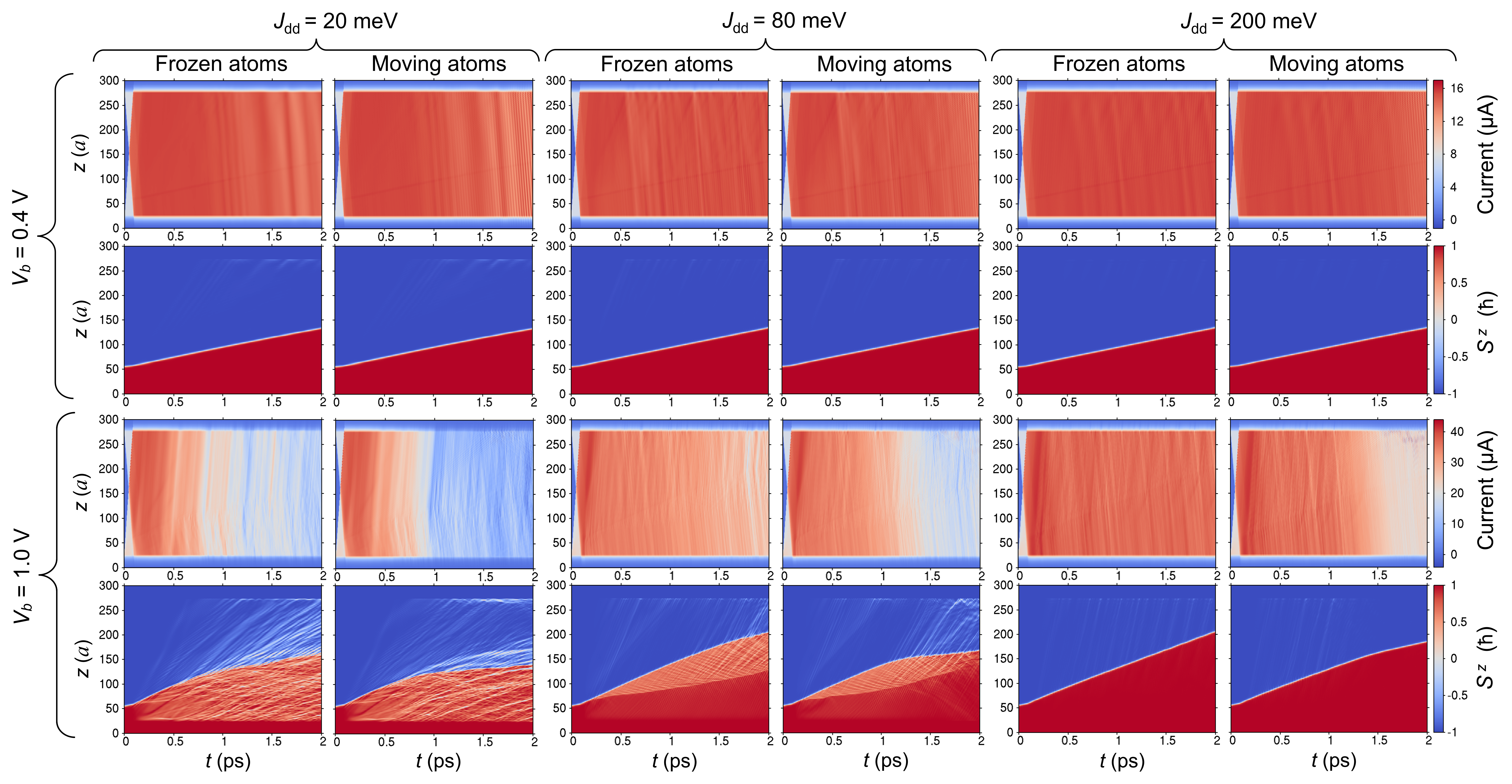}{Heatmaps in pairs of columns for each given value of $J_{dd}$ (with $J_z=J_{dd}/2$): 'frozen' and 'moving' atoms; and two pairs of rows showing the evolution of the bond currents and the $S^z$ component, at voltages given on the left.}{fig04}

These effects define a magnonic analogue of the phonon jets. Similarly to the phonons, the magnons are emitted preferentially into right-travelling states with a given crystal momentum $q$, this time around $\pi/4$ (see 2DFT panels for $V_b \leq 0.4$~V in Fig. \ref{fig03}). These can be related to two processes involving momentum exchange with the current-carrying electrons. As $\Abs{k_\mathrm{F}^{\uparrow}-k_\mathrm{F}^{\downarrow}}\approx \pi/4a$ (we are in a half-metallic regime with a minority spin wavevector approaching 0), where spin-up/down swap roles as majority/minority carriers across the DW, a majority spin traversing the wall non-adiabatically from left to right gives up crystal momentum of $q=\pi/4a$ as it becomes a minority spin on the far side. We see evidence of such non-adiabatic wall-crossing in Fig. \ref{fig01}(f). The second mechanism is spin-flip scattering, key to the magnon-drag effect \cite{magnons}. We see evidence for it in the suppression of the current in Figs. \ref{fig03} and \ref{fig04}, as well as the reduction of the DW speed below the adiabatic STT limit \cite{Tatara08}: at $V_b=1$~V the average $V_\mathrm{DW}\approx 89\,a/\mathrm{ps}$, while for the given average current of $I=34.0\,\upmu$A in the first ps after the transient, the adiabatic STT speed\cite{Tatara08} is $aPI/(2eS)=97\,a/\mathrm{ps}$, for a calculated spin-polarisation $P=91$~\% of the current. 

Between 1 and 2 V the DW velocity approaches the magnon group velocity (with an average of $300\,a/\mathrm{ps}$, see 2DFTs on Fig. \ref{fig03}). For such voltages the DW disintegrates and leaves behind a trail of magnetic structures coupled to electron excitations. Their signature in the 2DFT portraits is the continuum of states below the magnon dispersion -- strongly asymmetric in the direction of the electron flow.\footnote{Note that calculated magnon dispersions differ from a sine-square as per the nearest-neighbour Heisenberg model, because of the long-range $J_{dd}$ [see Eq. \ref{LLGel1}].} Therefore our DW cannot break the magnon speed barrier. \footnote{This limit also corresponds to the transition to strongly non-adiabatic interaction: $V_\mathrm{DW}>\omega_g w_\mathrm{DW} \approx 3.2 \times 10^4$~m/s, where $\omega_g \sim 0.05$ eV/$\hbar$ is the anisotropy-induced gap in our magnon dispersion.} The maximal DW velocity clearly remains below the speed of sound $c=190 a/\mathrm{ps}$. At artificially higher phononic stiffness and/or artificially small $J_{dd}$, it is also possible to enter a regime of strong phonon-magnon interaction, where the corresponding dispersions cross at a finite $q$-value. As this is difficult to realize in 1D magnetic structures, we just mention that the present findings still hold in this regime as the overall phonon-magnon interaction integrals remain small compared to the electron-phonon and electron-magnon ones. In other words, we do not observe supersonic nor 'super-magnonic' motion of the DW, like in the forementioned micromagnetic simulations \cite{Yan2013} and cannot reinstate the spin-Cherenkov effect for our choice of parameters. \footnote{This may not necessarily be the case for ferrimagnetic insulators, where DW velocities can be substantially higher.}

Lastly, Fig. \ref{fig04} examines the effect of the classical spin coupling $J_{dd}$ and shows combined lattice-magnetisation dynamics. At low $J_{dd}$, the magnon emission is more intense and blocks the DW motion early on due to the suppression of the current. This is suppressed at large $J_{dd}$ because of the increased magnetic stiffness. Comparing the currents for frozen atoms with those for the combined dynamics shows that the phonons contribute further to the suppression of the current and of the DW motion (without any quasi-particles $I\approx 40~\upmu$A at 1~V).

We thus witness not one but two superimposed mechanisms, whereby increasing bias unleashes a transition from ballistic transport to a regime where the incident electron quasi-momentum flux becomes almost entirely converted into a powerful phonon or magnon flux, or both, accompanied by a heavy suppression of the current. The phonon jet is associated with a clearcut threshold bias while in the magnon case the threshold is the breakdown of the DW as it approaches the magnon velocity.

This transition poses a fundamental limit to the current densities in electronic and spintronic devices. At the same time it opens up an mechanism for producing coherent monochromatic, or nearly monochromatic ($\Delta \omega_\mathrm{ph}/\omega_\mathrm{ph}\sim 15$~\%), phonon beams that may generate novel applications. Notice that a single-frequency phonon current requires small bias: Fig. \ref{fig02} shows how, at large bias, the violent dynamics splits the phonon frequency into a shower due to anharmonicity. Furthermore, the DW acts as a source of magnon current: at low bias the magnonic excitations in Fig. \ref{fig03} clearly originate from the DW with selected momenta due to the non-adiabatic spin propagation discussed earlier -- evidently, a coherent magnon source analogous to the phonon source above \footnote{There are a number of interesting questions around the dynamics of narrow DWs in finite 1D structures, bearing relation to the existence of Casimir forces acting on the DW, due to the wavelength constraints on the possible magnonic modes on either side. These will be the subject of a separate publication.} (see also SM\cite{SM} for further detail and discussion of calculated phononic and magnonic properties).

These predictions are our main finding. We now consider the experimental evidence. Above we assumed that $k_\mathrm{F} < \pi/2a$. If $k_\mathrm{F}  > \pi/2a$ then the emitted phonons with $q = 2 k_\mathrm{F}$ fall beyond the zone boundary and will manifest as backward-travelling phonons by Umklapp scattering. At the transition between these two cases, $k_\mathrm{F} = \pi/2a$, every emitted phonon lies on the zone boundary with $q = \pi/ a$ and equally qualifies as a back-travelling phonon with $q = -\pi/a$ which can be reabsorbed by the current-carrying electrons in the reverse process. Then we still expect a suppression of the current but no pile-up in the phonon population. Both of these situations we have seen in test simulations. 

The boundary case was observed experimentally in Au atomic wires \cite{rubio} and indeed long Au chains withstand high biases: 1 V or more. By contrast Pt, whose Fermi properties are dominated by a partially filled $d$-band \cite{pt}, shows very different behaviour and Pt chains fail at much lower biases \cite{jan}. We propose this as direct evidence for the phonon-jet effect, which shows that it is heavily material-dependent. Magnetic analogies could be sought at low temperatures and materials with heavy-magnon dispersions, opening up applications as magnon-current sources and techniques for the detection of DWs in magnetic nanostructures. These crucial dependencies on the electronic structure will, we hope, stimulate lively theoretical and experimental work to detect and exploit these effects.

\section {Acknowledgements}
We gratefully acknowledge funding from the Science Foundation Ireland: SFI Grants No. 18/SIRG/5515 and 18/NSFC/MANIAC. We thank the Irish Centre for High-End Computing (ICHEC) and the Trinity Research IT Centre (TCHPC) for the provision of computational facilities and support. We thank Andrew Horsfield, Myrta Gr{\" u}ning, Ray McQuaid and Stefano Sanvito for helpful discussions.

%\bibliography{references.bib}

%

\end{document}